\documentstyle[prl,aps,epsf]{revtex}

\tighten

\begin{document}
\draft

\twocolumn[\hsize\textwidth\columnwidth\hsize\csname @twocolumnfalse\endcsname

%
\title
{New model for surface fracture induced by dynamical stress}
%

\author
{J{\o}rgen Vitting Andersen$^{\ast}$ and Laurent J. Lewis$^\dagger$}

\address
{$^\ast$Department of Mathematics, Imperial College, Huxley Building, 180
Queen's Gate, London SW7 2BZ, England \\
$^\dagger$D\'epartement de physique et Groupe de recherche en physique et
technologie des couches minces (GCM), Universit\'e de Montr\'eal, Case
postale 6128, Succursale Centre-Ville, Montr\'eal, Qu\'ebec, Canada H3C 3J7}

\maketitle

\begin{abstract}

We introduce a model where an isotropic, dynamically-imposed stress induces
fracture in a thin film. Using molecular dynamics simulations, we study how
the integrated fragment distribution function depends on the rate of change
and magnitude of the imposed stress, as well as on temperature. A mean-field
argument shows that the system becomes unstable for a critical value of the
stress. We find a striking invariance of the distribution of fragments for
fixed ratio of temperature and rate of change of the stress; the interval
over which this invariance holds is determined by the force fluctuations at
the critical value of the stress.

\end{abstract}


\vskip2pc]


By experience most --- if not all --- materials will sooner or later develop
cracks. Yet, a profound understanding is largely missing of phenomena such as
how cracks initiate, the formation of networks of cracks and the resulting
distribution of fragments, the dynamics of crack propagation, and the
collective behavior of many interacting cracks. In this Letter we propose a
new model that addresses, at least in part, some of these questions. In the
model, an isotropic, dynamically-imposed stress, caused by material
properties changing in time, induces fracture in a surface material. The
problem is solved using molecular-dynamics simulations for a set of beads
interacting with one another via a continuous potential. This model should be
relevant to many phenomena that are known to lead to macroscopic fracture,
such as desiccation\cite{Groisman,Allain,Skjeltorp,abj} or
expansion\cite{Anifrani,asl}, changes in chemical composition\cite{Namgoong},
changes in temperature\cite{Becker}, or change of phase of the surface layer.

On the basis of a mean-field argument, we demonstrate that the system becomes
unstable for a critical value of the stress. We find a striking invariance of
the distribution of fragments for a fixed ratio of temperature and rate of
change of the stress; the interval over which this invariance holds is
determined by the force fluctuations at the critical value of the stress.

{\it Model} ---
We represent the thin film on a coarse-grained scale by beads that mutually
interact via a continuous potential which we take to be of the Lennard-Jones
form, $4 \epsilon [ (\sigma/r)^{12} - (\sigma/r)^6]$, where $r=|\vec{r}|$ is
the distance between two particles. An isotropic stress is imposed by having
$\sigma$ change in time ($t$), reflecting a change in the range
of the interactions on the surface \cite{note2}. For simplicity we limit
ourselves to the case where the material is initially unstressed and $\sigma
(t)$ decreases monotonically with time. This corresponds to a surface where
the induced stress makes the material rupture in a state of tension.

The dynamics of the beads obeys Newton's second equation, i.e., the system is
simulated using molecular dynamics (MD). We assume the surface layer to be in
contact with a heat bath at temperature $T$; this is done by periodically
rescaling the velocities to a fixed kinetic energy\cite{Tildesley}. The units
are chosen so that the mass $m= \epsilon \equiv 1$. In its initial
(stress-free) state, the surface layer consists of a triangular lattice with
lattice constant $a_0 = 2^{1\over 6} \sigma_0$, where $\sigma_0 =
\sigma(t_0)$. Periodic boundary conditions are used to eliminate surface
effects. The consequence of decreasing $\sigma (t)$ is to put all beads under
tensile stress, i.e., each bead feels attracted by its neighbors. We assume
$\sigma (t)$ to decrease linearly in time until it attains a final value
$\sigma_f$ at time $t_f$, whereafter it remains constant. An effective strain
parameter of the overlayer is defined by $s(t) \equiv [\sigma_0 -
\sigma(t)]/\sigma_0$. The rate of change (``speed'') of $\sigma$ is denoted
$v \equiv \partial \sigma(t)/\partial t$.

We are interested in the fracture pattern at $t=\infty$ which is obtained in
practice by choosing a large enough $t_f$, whose value depends on $s(t_f)$,
$v$, and $T$; the latter three parameters determine completely the fracture
pattern. In order to calculate the probability $P(f)$ for having a fragment
of size $f$, we discretize the system into cells of size $\sigma (t)$. A
fragment is then defined as a cluster of beads that are nearest or
next-nearest neighbors to one another.

As $\sigma$ changes with time, each bead will evolve from a position of
global energy minimum to a local minimum state. The local minimum energy
state is stable, however, for $\sigma (t)$ close to $\sigma_0$, since the
system would need instant cooperative motion of all the beads in order to
rearrange into the global minimum-energy state whose lattice parameter is
$a=2^{1 \over 6} \sigma (t)$. Due to the many body nature of the system, each
bead will see an energy landscape that changes as the positions of
neighboring beads change, and as $\sigma$ changes in time. The cooperative
motion of the beads create dynamical and spatial barriers between, on the one
hand, local metastable minimum energy states, and, on the other hand, the
global minimum energy state.

For increasing values of $\sigma$, the initial configuration eventually
become unstable. Neglecting fluctuations in the positions of the beads, each
will experience a mean field potential from its nearest neighbors given by:
   \begin{eqnarray}
   V(r,\sigma) &=&
   12 \epsilon [ ( {\sigma \over r} )^{12} - ({ \sigma \over r})^6
   + ( {\sigma \over 2a-r} )^{12} - ({ \sigma \over 2a-r})^6 ]. \nonumber
   \label{mfpot}
   \end{eqnarray}
We are interested in the behavior of $V(r,\sigma)$ at the point $r=a+\delta$
with $\delta$ small. Expanding the above to fourth order in $\delta$, we
find:
   \begin{eqnarray}
   V(\delta,\sigma) &=&
   12 \epsilon ({\sigma \over a})^{6} \{
   2 [({ \sigma \over a})^6 - 1]
   + [156 ({ \sigma \over a})^6 - 42]({ \delta \over a})^2 \nonumber \\
    & & + [32760 ({ \sigma \over a})^6 - 3024]({ \delta \over a})^4
   + O(({ \delta \over a})^6) \}. \nonumber
   \label{mfpotexp}
   \end{eqnarray}
Thus, for $\delta$ small, the potential seen by a bead changes from a
harmonic single-well to a double-well potential as $\sigma$ decreases. This
happens when $V^{''}(\delta,\sigma)|_{\delta=0}$ changes sign, that is for
$\sigma_c = (7/26)^{1/6}a_0 = (7/13)^{1/6} \sigma_0 \approx 0.90 \sigma_0$.
In general, the existence of a critical $\sigma_c$ for an arbitrary
interaction $V(r,\sigma)$ is equivalent to $V^{''}(r,\sigma)|_{r=a}=0$ having
a solution. As $\sigma (t)$ approaches $\sigma_c$ from below, one large
fluctuation eventually takes place bringing one of the beads close to it's
new local minimum energy position. A cascade of similar events then spreads
out from beads adjacent to that which first broke the configurational
symmetry. The extent of the propagation of this cascade of events, and the
subsequent fracturing of the system, depends, as we will see, on $s(t_f)$,
$T$, and $v$, as well as on the fluctuations of forces when $\sigma =
\sigma_c$.

{\it Results} ---
Figs.~1a-c show snapshots of one system for different values of stress but
fixed temperature and stress speed. Fig.~1a corresponds to a stress $\sigma
(t)$ slightly larger than $\sigma_c$. The very first cracks have appeared and
shortly after the system completely disintegrates into many pieces,
characterized by a macroscopic Young's modulus that goes to
0\cite{asl,Abraham}. This has happened in Fig.~1b. In Fig.~1c, we have the
final state of the system when the stress no longer varies in time. The
effect of varying the speed $v$ can be seen in Figs.~1d and 1e: here, the
initial conditions are the same as in Figs.~1a-c, but $v$ is 8 times smaller.
The stress in Fig.~1d is the same as in Fig.~1b; clearly, a smaller rate of
change of the stress gives the system longer time to respond so that the
positions of the beads are correlated over a longer distances and the cracks
are straighter. As a result, the fragments in the final configuration,
Fig.~1e, are larger than they are under a rapidly-varying stress (compare
Fig.~1c).

If $T=0$ the absence of thermal fluctuations would mean that the system
remains in its initial state and never breaks, despite the fact that the
energy difference between initial and stressed states increases as $\sigma
(t)$ decreases. For $T \neq 0$\cite{note1} and $v \rightarrow \infty$, on the
other hand, the rupture of the system is completely dominated by
fluctuations, in which case the probability density $P(f)$ for having a
fragment of given size $f$ is given by a binomial distribution $P(f) =
K_{(6,f)} ({1 \over 6})^f ({5 \over 6})^{6-f}$, since each of the 6 neighbors
of a given bead has probability ${1 \over 6}$ of forming a cluster with that
bead. For finite $(T,v)$, finally, the fracturing is determined by the
coherent motion of the $N$ beads. In Fig.~2a-b we show the cumulative
probability distribution $P_>(f)$ for a given $T$ and different $v$; as we
have seen above, the smaller the value of $v$, the larger the fragments. In
Fig.~2a, $s(t_f)=0.5$, whereas $s(t_f)=0.75$ in Fig.~2b. In order to
calculate $P_>(f)$, we have averaged over 200--500 $N=100$ systems with
different initial configurations, all at the same temperature $T$. (We chose
to use many small systems rather than few large ones in order to get better
statistics). Finite-size scaling of $P_>(f)$ is shown in the inset of
Fig.~2a, which allows us to extend our results (for the given $(T,v)$) to the
case $N \rightarrow \infty$. The lines are fits to a log-normal distribution;
clearly, the data suggest this form of $P_>(f)$ for large $v$. This is the
signature of a fracturing process that happens in a multiplicative
manner\cite{lognormal}, where a given piece at a random point breaks into two
pieces, which themselves randomly break into two other pieces, etc. For very
small $v$, $P_>(f)$ crosses over to a Heaviside theta function, since in this
case breakdown happens due to one large crack spanning the whole system. The
speed for which $P_>(f)$ can no longer be described by a log-normal
distribution depends on $T$ and $N$, and is due to finite size effects.

An instantaneous change in $\sigma$ means a change in both the magnitude and
the fluctuations of the forces. We find the system to respond in a {\em
qualitatively} different manner to changes in $\sigma$ depending if it is $<
\sigma_c$ or $> \sigma_c$. For a broad range of speeds $v$, we find the
average magnitude of the force on the beads, $F \equiv \sum_i^N |f_i|/N$, and
its fluctuation, $\delta F \equiv \sum_i^N \sqrt{f_i^2-F^2}/N$, to be {\em
independent} of $v$ for $\sigma (t) < \sigma_c$. We have also calculated the
characteristic length, $\xi (t)$, of the stress field $F[\vec{r}(t)]$ by
taking the first moment of the radial averaged structure factor $S(k,t)$. In
\cite{abj}, a coarsening phenomenon of $F[\vec{r}(t)]$ prior to the first
fracture was found to be crucial for the subsequent rupture of the system; in
the present model, we observe no time evolution of $\xi (t)$ for $\sigma (t)
< \sigma_c$, and $\xi (t) \simeq a$. (however, when the first macro cracks
appear, $\xi (t)$ increases dramatically). Therefore, the observed dependence
of $P_>(f)$ on $v$ must be due to the way the system responds to changes in
$\sigma$ {\em after} the $\sigma_c$ point has been passed.

Whether or not the system has time to counteract the imposed stress passed
$\sigma_c$ depends on the timescale over which changes in $\sigma$ take place
compared to the response time of the system; the latter is determined by the
random thermal motion, i.e., kinetic energy $E_k$, of the beads. The ratio of
these two timescales is thus given by $\kappa \equiv \sigma v^{-1}/ (m^{1
\over 2} \sigma /E_k^{1/2}) = \sqrt{E_k}/v$. One therefore expect systems
with the same value of $\kappa$ to fracture in the same way. The fracture is
expected to be dominated by fluctuations for $\kappa \ll 1$, whereas for
$\kappa \gg 1$ it will have time to respond to the changing stress in a
correlated manner. This is in fact verified in Fig.~3 which shows a
remarkable invariance of $P_>(f)$ over almost 3 decades in temperature for
systems with two different values of $\kappa$. The lowest and highest
temperature in Fig.~3 for which the invariance of $P_>(f)$ no longer holds,
and the subtle temperature dependence at intermediate values, can be
understood from the dependence on stress of the force fluctuations $\delta
F$, shown in Fig.~4 for the same values of $T$ as Fig.~3. Because of
fluctuations, different temperatures lead to a critical $s_c$ (defined as the
$s$ for which $\delta F$ has its minimum) slightly different from the mean
field value of $s_c=(\sigma_0-\sigma_c)/\sigma_0=0.10$. The small temperature
dependence of $P_>(f)$ at intermediate temperatures can then be understood in
terms of a slight increase of $\delta F(s_c)$ with $T$, since one would
expect larger force fluctuations at $s_c$ to lead to smaller fragments. As
seen in Fig.~4, the only exception to this is the case of the highest $T$
where, on the contrary, a large $\delta F(s_c)$ leads to a large-fragment
tail in $P_>(f)$. The reason for this is that $T$ is so high that coalescence
of already-formed fragments takes place; coalescence is not observed for
lower $T$. Finally one also notes from Fig.~3 that deviations in $P_>(f)$
occur for very low temperatures, where the simple scaling argument leading to
invariance of $P_>(f)$ under a given $\kappa$ apparently no longer holds.

{\it Conclusion --- }
We have introduced a model where a dynamically-imposed stress induces
fracture in a thin film. Using molecular-dynamics simulations, we have shown
the accumulated fragment distribution function to obey a log-normal
distribution characteristic of fracturing processes which happen in a random
multiplicative manner. A mean field argument shows how the system undergoes
an instability for a critical value of the imposed stress. We find a striking
invariance of the fragment distribution function for a given ratio of
temperature and speed of stress; the interval over which this invariance
holds, is determined by the force fluctuations at the critical value of the
stress.

J.V.A.\ wishes to acknowledge support from the European Union Human Capital
and Mobility Program contract number ERBCHBGCT920041 under the direction of
Prof.\ E.\ Aifantis, as well as the hospitality of the D\'epartement de
Physique de l'Universit\'e de Montr\'eal, where part of this work was carried
out. This work was also supported by grants from the Natural Sciences and
Engineering Research Council of Canada and the ``Fonds pour la formation de
chercheurs et l'aide {\`a} la recherche'' of the Province of Qu{\'e}bec.

\vspace{-0.5cm}


\newpage
\widetext
\onecolumn
\begin{figure}
\caption{
Snapshots of a $N=1600$ system at different times $t$, with different change
of strain rate $v$ and different final strain $s(t_f)$. The initial
configuration is the same, and $T= 6.25 \times 10^{-5}$, in all cases. (a)
$s(t)=0.14, v=0.0125$, (b) $s(t)=0.25, v=0.0125$, (c) $s(t=t_f)=0.5,
v=0.0125$, (d) $s(t)=0.25, v=0.0015625$ and (e) $s(t=t_f)=0.5, v=0.0015625$.
}
\label{fig1}
\end{figure}
\begin{figure}
\caption{
Cumulative probability distribution $P_>(f)$ for finding a given fragment of
area larger than $f$, and $T= 6.25 \times 10^{-5}$ (a) $s(t_f)=0.5$ ; $v=$
0.025 ($\diamond$), 0.0125 (+), 0.00625 ($\Box$), 0.0042 ($\times$), 0.003125
($\triangle$) and 0.0015625 ($*$). The lines to are fits to a log-normal
distribution. Inset: finite size scaling with $s(t_f)=0.5$, $v=0.0125$; $N=$
100 ($\times$), 400 ($\triangle$) and 900 ($*$); $s(t_f)=0.75$, $v=$ 0.0125;
$N=$ 100 ($\diamond$), 400 ($+$) and 900 ($\Box$). (b) $s(t_f)=0.75$ ; $v=$
0.0375 ($\diamond$), 0.01875 ($+$), 0.009375 ($\Box$), 0.0046875 ($\times$),
0.003125 ($\triangle$), 0.002679 ($*$) and 0.002344 (small white circle). The
lines to are fits to a log-normal distribution.
}
\label{fig2a}
\end{figure}
\setcounter{figure}{1}
\begin{figure}
\caption{
Inset to Fig.~2a.
}
\label{fig2a_inset}
\end{figure}
\setcounter{figure}{1}
\begin{figure}
\caption{$ $Fig.2b}
\label{fig2b}
\end{figure}

\begin{figure}
\caption{
$P_>(f)$ versus $f$ for fixed value of $\kappa \equiv E_k^{1/2} / v$ and
$s(t_f)$ for a $N=100$ system. (i) $\kappa$ = 1.03, $s(t_f)=0.75$, and
$(T,v)=$ ($6.4 \times 10^{-2}$, 0.30) ($\diamond$), ($1.6 \times 10^{-2}$,
0.15) (+), ($4 \times 10^{-3}$, 0.075) ($\Box$), ($10^{-3}$, 0.0375)
($\times$), (2.5 $\times 10^{-4}$, 0.01875) ($\triangle$) and (6.25 $\times
10^{-5}$, 0.009375) (*). (ii) $\kappa$ = 1.55, $s(t_f)=0.5$, and $(T,v)=$
($6.4 \times 10^{-2}$, 0.20) (large black circle), ($1.6 \times 10^{-2}$,
0.10) (black circle), ($4 \times 10^{-3}$, 0.05) (small white circle),
(10$^{-3}$, 0.025) (white circle), (2.5 $\times 10^{-4}$, 0.0125) (large
white circle) and (6.25 $\times 10^{-5}$, 0.00625) (small black circle).
}
\label{fig3}
\end{figure}

\setcounter{figure}{3}
\begin{figure}
\caption{
$\delta F$ versus $s$ for $\kappa =1.55$ and $(T,v)=$ ($6.4 \times 10^{-3}$,
0.20) ($\diamond$), ($1.6 \times 10^{-2}$, 0.10) (+), ($4 \times 10^{-3}$,
0.05) ($\Box$), (10$^{-3}$, 0.025) ($\times$), (2.5 $\times 10^{-4}$, 0.0125)
($\triangle$) and (6.25 $\times 10^{-5}$, 0.00625) (*).
}
\label{fig4}
\end{figure}
\setcounter{figure}{3}
\begin{figure}
\caption{Fig.4-inset}
\label{fig4_inset}
\end{figure}


\begin{references}
\bibitem{note1}
We choose in all cases $T \ll T_c$, where $T_c$ is the temperature for the
solid-liquid phase transition of the overlayer.
\end{references}
\end{document}